\RequirePackage{booktabs} % needed for cline
\documentclass[journal=jacsat]{achemso}

\RequirePackage{hyperref}
\hypersetup{
        colorlinks,
        breaklinks=true,
        plainpages=false,
        citecolor=blue,
        linkcolor=blue,
        urlcolor=blue,
        bookmarksopen=true,
        bookmarksnumbered=false,
        bookmarksdepth=5
}
\usepackage{achemso}
\usepackage{graphicx}
\graphicspath{{images/}}
\usepackage{setspace}
\usepackage{array}
\usepackage{makecell}
\usepackage{multirow}
\usepackage{float}
\usepackage{xcolor}
\usepackage{geometry}
\usepackage{svg}
\usepackage{anyfontsize}
\usepackage{tablefootnote}

\usepackage{amsmath,amssymb,amsfonts}%
\usepackage{amsthm}%
\usepackage{mathrsfs}%
\usepackage[title]{appendix}%
\usepackage{textcomp}%
\usepackage{manyfoot}%
\usepackage{booktabs}%
\usepackage{algorithm}%
\usepackage{algorithmicx}%
\usepackage{algpseudocode}%
\usepackage{listings}%

\newcommand*\secref[1]{Section~``\nameref{#1}''}

\usepackage{glossaries}
\makeglossaries
\newacronym{vfs}{VFS}{virtual forward synthesis}
\newacronym{smiles}{SMILES}{simplified molecular-input line-entry system}
\newacronym{api}{API}{application programming interface}
\newacronym{json}{JSON}{JavaScript object notation}
\newacronym{smarts}{SMARTS}{SMILES arbitrary target specification}
\newacronym{cas}{CAS}{Chemical Abstracts Service}
\newacronym{calb}{CALB}{Candida antarctica lipase B}
\newacronym{ps}{PS}{polystyrene}
\newacronym{umap}{UMAP}{Uniform Manifold Approximation and Projection}
\newacronym{tg}{T$_g$}{glass transition temperature}
\newacronym{tm}{T$_m$}{melting temperature}
\newacronym{td}{T$_d$}{decomposition temperature}
\newacronym{E}{E}{Young's modulus}
\newacronym{tsb}{$\sigma_b$}{tensile strength at break}
\newacronym{epsb}{$\epsilon_b$}{elongation at break}
\newacronym{tc}{T$_c$}{ceiling temperature}
\newacronym{dh}{$\Delta H$}{enthalpy of polymerization}
\newacronym{ds}{$\Delta S$}{entropy of polymerization}
\newacronym{cp}{C$_p$}{heat capacity}
\newacronym{rop}{ROP}{ring-opening polymerization}
\newacronym{ml}{ML}{machine learning}
\newacronym{gpr}{GPR}{Gaussian process regression}
\newacronym{mtnn}{MTNN}{multitask neural network}
\newacronym{dft}{DFT}{density functional theory}
\newacronym{sql}{SQL}{structured query language}
\newacronym{sas}{SAScore}{synthetic accessibility score}
\newacronym{rmse}{RMSE}{root mean squared error}
\newacronym{md}{MD}{molecular dynamics}

\title{An Informatics Framework for the Design of Sustainable, Chemically Recyclable, Synthetically-Accessible and Durable Polymers}

\author{Joseph Kern}
\affiliation{School of Materials Science and Engineering, College of Engineering, Georgia Institute of Technology, 771 Ferst Dr. N.W., Atlanta, GA 30318, U.S.A.}

\author{Yongliang Su}
\affiliation{School of Chemistry \& Biochemistry, College of Sciences, Georgia Institute of Technology, 901 Atlantic Drive NW, Atlanta, GA 30318, U.S.A.}

\author{Will Gutekunst}
\affiliation{School of Chemistry \& Biochemistry, College of Sciences, Georgia Institute of Technology, 901 Atlantic Drive NW, Atlanta, GA 30318, U.S.A.}

\author{Rampi Ramprasad}
\email{rampi.ramprasad@mse.gatech.edu}
\affiliation{School of Materials Science and Engineering, College of Engineering, Georgia Institute of Technology, 771 Ferst Dr. N.W., Atlanta, GA 30318, U.S.A.}

\begin{document}

\abstract{We present a novel approach to design durable and chemically recyclable \gls*{rop} class polymers. This approach employs digital reactions using \gls*{vfs} to generate over 7 million ROP polymers and machine learning techniques to rapidly predict thermal, thermodynamic and mechanical properties crucial for application-specific performance and recyclability. This combined methodology enables the generation and evaluation of millions of hypothetical ROP polymers from known and commercially available molecules, guiding the selection of approximately 35,000 candidates with optimal features for sustainability and practical utility. Three of these recommended candidates have passed validation tests in the physical lab — two of the three by others, as published previously elsewhere, and one of them is a new thiocane polymer synthesized, tested and reported here. This paper presents the framework, methodology, and initial findings of our study, highlighting the potential of \gls*{vfs} and machine learning to enable a large-scale search of the polymer universe and advance the development of recyclable and environmentally benign polymers.}

\maketitle

% =========================================================================== %
% Intro
\section{Introduction}

Plastics, central to everyday life and pivotal in diverse applications ranging from food packaging to electronic components, cause a concerning amount of environmental pollution. Studies reveal pervasive microplastic contamination globally that negatively impact humans, plants, and animals \cite{ozbayDesignOperationEffective2021, shahulhamidWorldwideDistributionAbundance2018, barbozaMarineMicroplasticDebris2018, campanaleDetailedReviewStudy2020,geyerProductionUseFate2017a}. The quest for sustainable alternatives that balance the beneficial attributes of plastics (such as cost-effectiveness, durability, and performance) with environmental considerations (such as recyclability, and reduced ecological footprint) is a significant focus in contemporary materials development.

Creation of new plastics amenable to chemical recycling, i.e., transforming them back to monomers at the end of their life, will be enormously beneficial. Traditional mechanical recycling methods suffer from degradation limits \cite{schynsMechanicalRecyclingPackaging2021}, whereas chemical recycling promises near-infinite recyclability. However, polymers that can undergo chemical recycling must still meet the demands of their application needs. A new take-out container that is recyclable sounds attractive, but consumers will not use it if it breaks after holding just a single item or if it is too expensive.

\begin{figure}
\centering
\includegraphics[width=\textwidth]{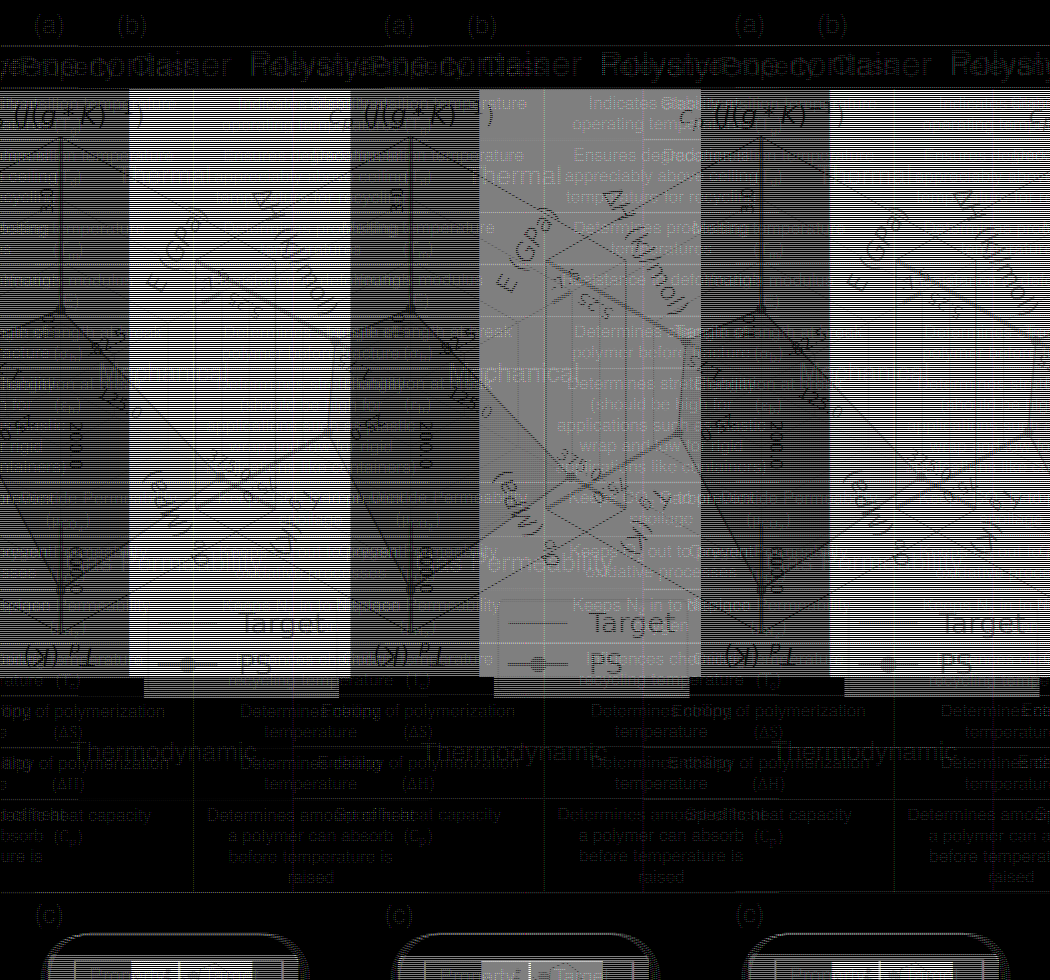}
\caption{(a) Classification of key polymer properties, including specific properties within each class, and their relevance. (b) Radar chart illustrating the design targets for a polystyrene container, highlighting polystyrene properties in blue and target values in orange. (c) Overview of the material informatics workflow, starting with the definition of screening criteria, followed by the development of \gls*{ml} models, defining of the polymer search space being explored, and ending with the screening and recommendation of suitable candidates for further experimental validation and testing.}
\label{fig:fig1}
\end{figure}

As depicted in \autoref{fig:fig1} (a), polymer design can be complex. Various classes of properties must be considered as they profoundly influence the polymer's performance within consumer and industrial environments. These encompass thermal properties, such as the \gls*{tg} and \gls*{tm}, which not only dictate stability at operational temperatures but also affect processing conditions. Additionally, mechanical properties like \gls*{E}, \gls*{tsb}, and \gls*{epsb} play crucial roles in determining the polymer's stiffness, strength, and stretchability. Certain applications, such as take-out containers for food demand thermal insulation properties to prevent heat transfer or minimize the risk of burn injuries upon contact. These requirements necessitate considerations such as gas permeability, while others necessitate considerations such as gas permeability. Adding thermodynamic properties such as the \gls*{dh} and \gls*{tc}, essential to assess chemical recyclability, complicates the design process further.

In this work, our focus is on the design of suitable chemically recyclable polymers for a specific application, namely, a replacement material for \gls*{ps} used in containers. This focus is significant considering \gls*{ps}'s notable presence in plastic production, its recyclability challenges, and its associated environmental and health concerns. \gls*{ps} constituted 6.8\% of plastic production in Europe in 2019 \cite{kuennethBioplasticDesignUsing2022}, while the U.S. alone saw the creation of 220 thousand tons of \gls*{ps} containers, bags, sacks, and wraps in 2018 \cite{FrequentQuestionsRegarding2018}. Such substantial volumes are required for recyclable alternatives, as economies of scale often dictate affordability \cite{ellenmacarthurfoundationNewPlasticsEconomy2017}. While \gls*{ps} is technically recyclable, it is not commonly recycled due to prohibitive costs \cite{miaoCurrentTechnologiesDepolymerization2021, ellenmacarthurfoundationGlobalCommitment20212021}. Styrene, the monomer used in its production, is also classified as ``reasonably anticipated to be a human carcinogen" by the Department of Health and Human Services (DHHS) and the National Toxicology Program (NTP) \cite{StyrenePublicHealth}. \gls*{ps} microplastics have also been identified as potential immune system stimulants and toxic to freshwater organisms \cite{nugnesToxicImpactPolystyrene2022, hwangPotentialToxicityPolystyrene2020}.

In \autoref{fig:fig1} (b), a radar chart illustrating the properties of \gls*{ps} (depicted as blue dots) alongside the specific property targets set for our design (represented by orange lines) is presented. These targets, explicitly outlined in the first box of the informatics workflow of \autoref{fig:fig1} (c), were carefully selected through a comprehensive analysis of \gls*{ps} properties, coupled with considerations of the typical operating conditions of a container.

Concerning thermal properties, our design prioritizes a \gls*{tg} value surpassing the boiling point of water (373 K), ensuring the container's integrity under operational conditions typical of \gls*{ps} containers. Additionally, maintaining the polymer in a glassy state below this temperature is essential for structural rigidity. A \gls*{td} set 100 K above the boiling point of water is chosen to prevent decomposition during thermally-induced, chemical recycling.

Addressing mechanical considerations, we establish a \gls*{tsb} exceeding 39 MPa to ensure the container's resistance to breakage when subjected to typical loads. Similarly, a minimum \gls*{E} exceeding 2 GPa is stipulated to mitigate excessive bending when loaded with contents, aligning closely with the properties exhibited by PS \cite{OverviewMaterialsPolystyrene}.

In assessing thermodynamic attributes, a \gls*{cp} akin to that of \gls*{ps} is desired, emphasizing the necessity for thermal insulation to prevent burns from hot contents. While considerations such as thermal conductivity are desirable, rapid and accurate models to predict this property for polymers are not available. Unlike other properties where exceeding a threshold is sought, the \gls*{dh} is constrained within a narrow range of -10 to -20 kJ/mol. It's noteworthy that the true quantity of interest that determines the polymerization/depolymerization equilibrium is the \gls*{tc}, which is defined as the ratio of \gls*{dh} to the \gls*{ds}. If \gls*{dh} is too negative, the polymer may not be depolymerizable, while if it's too high, it may not be polymerizable.

For each of these properties, we developed rapid and accurate \gls*{ml} models to predict them from a polymer's molecular structure, as displayed in the informatics workflow's second box in \autoref{fig:fig1} (c)\cite{kuennethPolyBERTChemicalLanguage2023,tolandAcceleratedSchemePredict2023}. These models offer a significant advantage over physics-based simulation techniques such as \gls*{dft} and classical \gls*{md} due to their remarkable speed, enabling them to evaluate the millions of polymers in our search space efficiently. 

Notably, the modeling of polymer recyclability remains an emerging field. As such our exploration, illustrated by the sample \gls*{umap} in the third box of \autoref{fig:fig1} (c), concentrated on the search space of \gls*{rop} polymers.  We chose to narrow our focus to \gls*{rop} polymers because of their demonstrated potential to meet the necessary thermodynamic criteria for facilitating chemical recycling to monomers \cite{coatesChemicalRecyclingMonomer2020a, zhangRecyclablePolyesterLibrary2023}. Moreover, \gls*{rop} polymers have garnered significant attention in the pharmaceutical industry owing to their customizable properties, biocompatibility, and biodegradability \cite{phanFunctionalInitiatorsRing2020, tianBiodegradableSyntheticPolymers2012, capassopalmieroStrategiesCombineROP2018}. Our \gls*{dh} model is specifically trained on a dataset tailored to \gls*{rop}, thereby improving its accuracy in predicting outcomes related to \gls*{rop} polymers \cite{tolandAcceleratedSchemePredict2023, tranRecyclablePolymersRingOpening2022}. 

Subsequently, after a thorough exploration of the molecular space to find candidates that could undergo \gls*{rop}, we screened over seven million hypothetical polymer designs (encompassing 9 polymer classes) to identify promising candidates that met our stringent screening criteria. The screened candidates are then recommended for further in depth lab-based studies, as depicted in the final box of \autoref{fig:fig1} (d). Given the vastness of this search space, traditional Edisonian trial-and-error methods would prove impractical. Consequently, there is a pressing need for computational techniques capable of swiftly predicting polymer properties and identifying promising candidates, as has been done here. In the subsequent sections, we delineate our polymer design process and spotlight designs deemed most promising for environmentally friendly alternatives to \gls*{ps}. We also highlight three experimentally validated designs, two of which have been previously published by others, and a third, a novel thiocane polymer, which we have synthesized, tested, and report on here for the first time.

% =========================================================================== %
% Results

\section{Results}
\label{results}
\subsection{Virtual Forward Synthesis (VFS) Reactions}

\begin{figure}
\centering
\includegraphics[width=\textwidth]{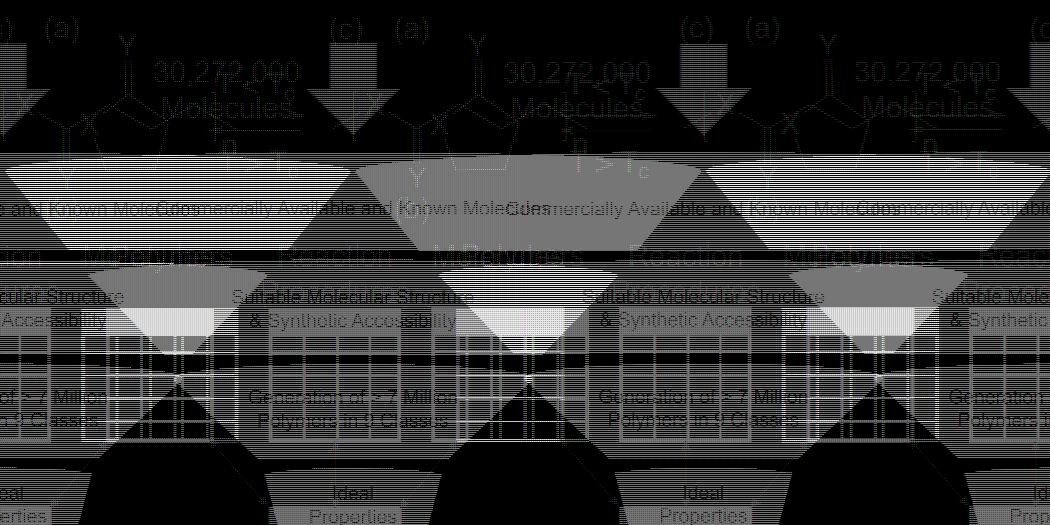}
\caption{(a) Schematic illustrating the (de)polymerization process of ring-opening polymers. Above the ceiling temperature, the polymer should undergo depolymerization, while maintaining stability below this threshold. (b) A simplified representation of the PostgreSQL database schema. Molecules, the reactions they go through, and the polymers they create are stored in separate tables and joined together via a central mapping table. (c) Funnel diagram demonstrating the down-selection process for promising polymers.}
\label{fig:fig2}
\end{figure}

As stated previously, in the pursuit of our design goals, we narrowed our focus to \gls*{rop} polymers, because of their potential to meet the requisite thermodynamic criteria for facilitating chemical recycling to monomers \cite{coatesChemicalRecyclingMonomer2020a, zhangRecyclablePolyesterLibrary2023}. As depicted in \autoref{fig:fig2} (a), the polymer should undergo depolymerization into its monomer ring form above its \gls*{tc}, while remaining stable in polymer form below it.

To initiate the search for viable polymer candidates, we employ a \gls*{vfs} approach. This method involves the systematic generation of hypothetical polymers from a database of initial monomers, following established reaction pathways. Recent advancements in this area, exemplified by initiatives such as the Open Macromolecular Genome and SMiPoly \cite{kimOpenMacromolecularGenome2023, ohnoSMiPolyGenerationSynthesizable2023}, have underscored the potential of this methodology. Unlike these predecessors, our approach places a significant emphasis on database design, as detailed in \secref{methods:db}, and supports the integration of multi-step reaction pathways.

The underlying structure of our database schema is illustrated in \autoref{fig:fig2} (b), while (c) depicts the procedural workflow of the \gls*{vfs} technique. Within our database, a \gls*{sql} table houses the \gls*{smiles} of both known and hypothetical molecules. During each \gls*{vfs} reaction procedure, we search this repository for known molecules possessing the required substructures via \gls*{smarts} substructure queries \cite{DaylightTheorySMARTS}. Subsequently, we employ a stringent selection criterion, excluding excessively intricate molecules (those with \gls*{sas} $>$ 7) \cite{ertlEstimationSyntheticAccessibility2009a}.

Following this filtration process, we subject the retained molecules to the reaction procedure to generate hypothetical polymers. Additionally, for multi-step reactions, we generate hypothetical monomers, essential for \gls*{dh} predictions, and store them in the molecules table if they are new. Subsequently, both molecules and polymers are converted to \gls*{smiles} format and subjected to canonicalization, ensuring that the same string is output for the same structure \cite{weiningerSMILESChemicalLanguage1988}. This canonicalization step is crucial for database querying efficiency. While RDKit's CanonSmiles function is used for molecule canonicalization \cite{RDKit}, the canonicalize\_psmiles package \cite{kuennethPolyBERTChemicalLanguage2023,kuennethRamprasadGroupCanonicalize_psmiles2024} serves this purpose for polymers.

The canonicalization of \gls*{smiles} strings enables efficient querying within the database, operating at O(log(n)) complexity due to B-Tree indexing on the canonical\_smiles column. Upon identification or generation of IDs, a mappings table is used to store the pertinent associations among molecule ID, reaction procedure ID, and polymer ID. This facilitates the swift retrieval of reactants necessary for promising polymer synthesis, alongside the proposed reaction pathway. The identification of promising polymers entails leveraging \gls*{ml} techniques to predict their properties and subsequently evaluating them against predefined screening criteria, as identified in \autoref{fig:fig1} (c).

In \autoref{tab:tab1}, a simplified version of the \gls*{rop} reaction is displayed, along with the class of monomer ring being opened, the number of polymers generated from each class and how many successfully met all target criteria. Our investigation encompasses ring-opening reactions of ethers, thioethers, esters, thioesters, thionoesters, amides, cycloalkenes, carbonates, and thiocanes. For thioethers, thioesters, and thionoesters, in addition to exploring commercially available options, we also explored hypothetical monomer designs by swapping the appropriate oxygen in an ether or esters with a sulfur. 

Similarly, for thiocanes, because so few commercially available molecules with the thiocane structure existed, we developed a two step reaction procedure guided by the expertise of our polymer chemists. This two step procedure had two variants, R$_{1}$ and R$_{2}$ as outlined in \autoref{tab:tab1}. The variants differ in their functionalization step: R$_1$ involves modifying the vinyl sulfide's vinyl component with a terminal alkyne-containing molecule, whereas R$_2$ involves functionalizing the ring's ketone component with a bromine-containing molecule. Notably, both the alkyne and bromine groups are eliminated during this initial reaction step, leaving behind the R-group attached to the thiocane. The resulting thiocane can be either the R$_1$ or R$_2$ variant, but not both. Both variants undergo ring-opening in the second step, yielding a novel polythiocane. Detailed multi-step reaction procedures, inclusive of the \gls*{smarts} employed for search and multi-step reaction \gls*{smarts}, are outlined in the supplementary ``reaction\_procedures.json'' file.

\begin{table}
\caption{ROP Virtual Synthesis Screening Results\tablefootnote{With the exception of thiocanes, the inclusion of `R' in the reaction signifies that the ring can be of any size, and the `R' ring atoms can be any element with any side chain attached.  For thiocanes, `R$_1$' and `R$_2$' involve distinct modifications to the molecule, with `R$_1$' adding a terminal alkyne containing molecule to the vinyl component and `R$_2$' functionalizing the ketone component with a bromine-containing molecule, resulting in either the `R$_1$' or `R$_2$' thiocane variant after the elimination of the alkyne and bromine groups.}}
\label{tab:tab1}

\begin{tabular}{@{}m{4cm}ccc@{}}
\toprule

\centering \makecell{Reaction \\ (X = S, O)} & \makecell{Monomer \\ Class} & \makecell{Polymers \\ Generated} & \makecell{Promising \\ Candidates} \\

\midrule

\multirow{2}{*}[-0.1cm]{\includegraphics[width=4cm]{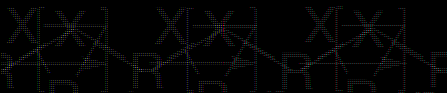}} & Thioether & 2,797,024 & \makecell{12,100 \\ (0.43\%)} \\

\cline{2-4}

& Ether & 2,321,545 & \makecell{8,850 \\ (0.38\%)} \\

\hline

\includegraphics[width=4cm]{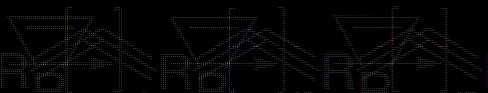} & Cycloalkene & 1,164,917 & \makecell{2,876 \\ (0.25\%)} \\

\hline

\includegraphics[width=4cm]{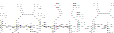} & Thiocane & 306,177 & \makecell{126 \\ (0.04\%)} \\

\hline

\includegraphics[width=4cm]{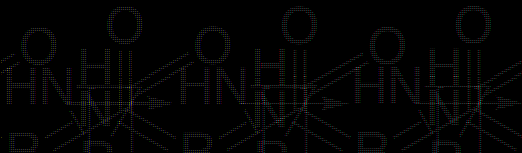} & Amide & 256,223 & \makecell{6,827 \\ (2.66\%)} \\

\hline

\multirow{3}{*}[-0.5cm]{\includegraphics[width=4cm]{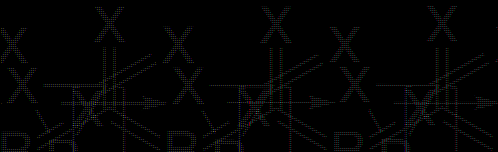}} & Thioester & 225,649 & \makecell{1,177 \\ (0.52\%)} \\

\cline{2-4}

& Thionoester & 130,230 & \makecell{ 1,777 \\ (1.36\%)} \\

\cline{2-4}

& Ester & 80,607 & \makecell{1,578 \\ (1.96\%)} \\

\hline

\includegraphics[width=4cm]{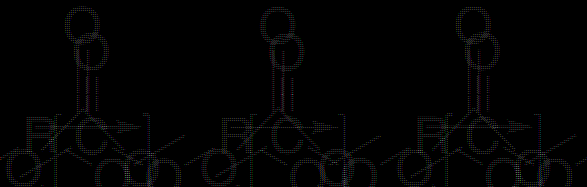} & Carbonate & 821 & \makecell{10 \\ (1.22\%)} \\

\hline

& Total & 7,301,681 & \makecell{35,321 \\ (0.48\%)} \\

\bottomrule
\end{tabular}
\end{table}

\subsection{Predictive Models}

\begin{figure}
\centering
\includegraphics[width=\textwidth]{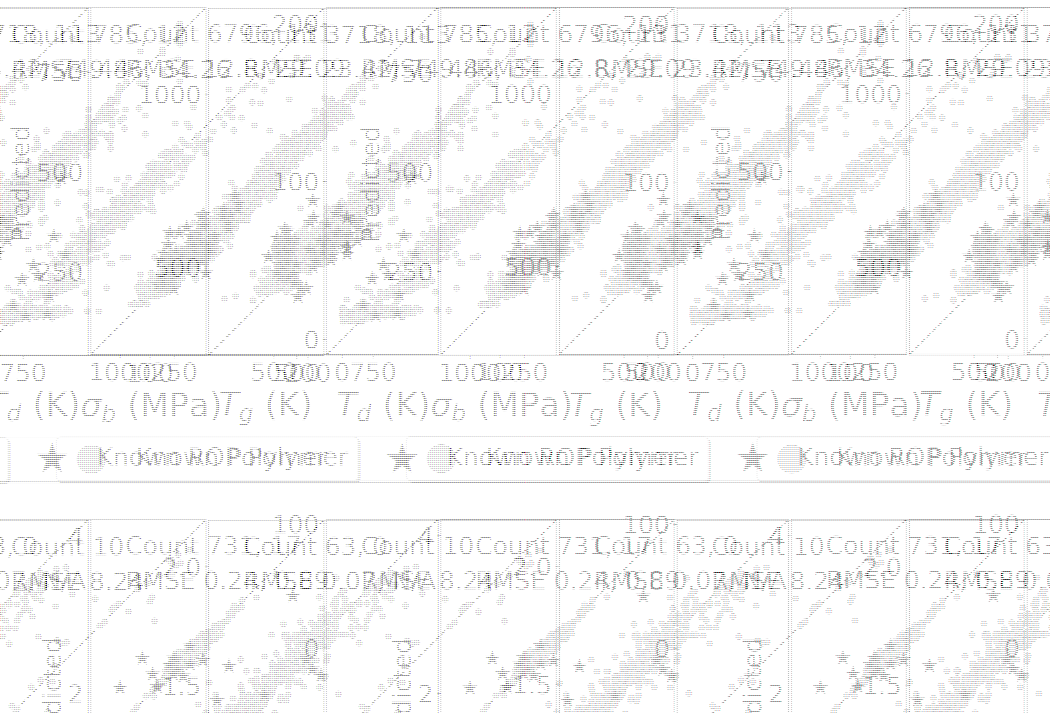}
\caption{Parity plot illustrating model performance across different properties. The first set of values in the top left showcase counts and RMSE on the model test data of known polymers, which are represented by red circles. The second set delineates the dataset size and model performance specifically on known \gls*{rop} polymers, denoted by black stars. For enthalpy of polymerization, dots depict training data, while black stars indicate test data, as the model was exclusively trained on ROP polymers.}
\label{fig:fig3}
\end{figure}

Polymers generated in this study had their properties predicted using two subsets of previously developed and published models \cite{tolandAcceleratedSchemePredict2023,kuennethPolyBERTChemicalLanguage2023}, as described in \secref{methods:models}: a \gls*{gpr} model to predict \gls*{dh}  based on the polymer and monomer\cite{tolandAcceleratedSchemePredict2023}, and a \gls*{mtnn} trained on homo and copolymer data to predict all other properties\cite{kuennethPolyBERTChemicalLanguage2023}. Parity plots illustrating the \gls*{mtnn} models' performance on test data of known polymers (red dots) and our dataset of known \gls*{rop} polymers (black stars) are depicted in \autoref{fig:fig3}. The count and \gls*{rmse} shown represents firstly the test dataset size and model performance on the test data, then on the known \gls*{rop} polymers. For \gls*{dh}, the black stars and labeled \gls*{rmse} and count indicate the performance on test \gls*{rop} data and the size of the test dataset, while the colored circles represent the training data.

With the exception of \gls*{dh}, the models generally perform worse for known \gls*{rop} polymers than their known test dataset chemistries. This outcome is unsurprising given the limited variety of \gls*{rop} chemistries in the original training data. Of the 138 unique \gls*{rop} polymers with thermal property data available, only eight were seen in the training data, and of the twenty with mechanical property data available, none were in the training data. However, predictions for thermal properties tend to align closely with the parity line, indicating acceptable performance. In contrast, for mechanical properties, the models tend to over-predict performance. In addition to the lack of \gls*{rop} data, this discrepancy could be attributed to the training process, as the models only consider molecular structure, while factors such as molecular weight, which contribute significantly to mechanical properties, are not considered. It is possible that while the \gls*{rop} molecular structure enhances strength, the molecular weight remains insufficient \cite{camaSyntheticBiodegradableMedical2016}.

This observation is supported by two outlier polymers in the \gls*{tsb} plot. These outliers exhibit predicted values of 58 and 77 MPa, whereas their actual values are 3.4 and 3.1 MPa, respectively. Notably, the molecular weights associated with these polymers are very low at 18 and 10.8 kDa, compared to other \gls*{rop} polymers in the dataset, which range from 109-126 kDa (44 MPa), 198 kDa (39 MPa), to 266-438 kDa (46.4 MPa) \cite{zhangRecyclablePolyesterLibrary2023a, satheOlefinMetathesisbasedChemically2021a, tuBiobasedHighPerformanceAromatic2021}.

Similarly, in the case of \gls*{E}, one outlier exhibits a predicted value of 2.07 GPa, while its true value is 0.17 GPa, with a low molecular weight of 18 kDa \cite{zhangRecyclablePolyesterLibrary2023a}. However, this theory fails to account for some other inaccuracies of the model, as evidenced by two other polymers with predicted values of 2.44 and 2.08 GPa. Their true values were considerably lower at 0.45 and 0.64 GPa, despite higher molecular weights of 85.6 kDa and 69.3 kDa.\cite{liToughThermallyRecyclable2021, tuBiobasedHighPerformanceAromatic2021}. These instances suggest that the model has not encountered enough similar chemistry.

\begin{figure}
\centering
\includegraphics[width=\textwidth]{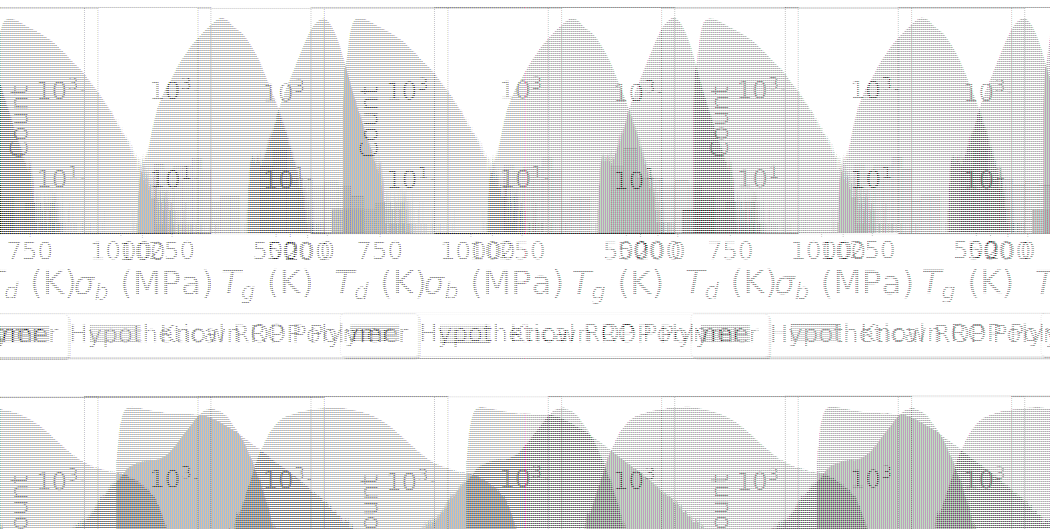}
\caption{Histograms illustrating property distributions for hypothetical ROP polymers (green) and known ROP polymers (black).}
\label{fig:fig4}
\end{figure}

\autoref{fig:fig4} shows the histograms representing the distribution of the predicted polymer properties for the entire pool of over seven million hypothetical ROP polymers generated here (green), as well as for the properties of presently known \gls*{rop} polymers (black). Note that the y-axis is in log scale. Based on \autoref{fig:fig4}, it can be concluded that existing \gls*{rop} polymers generally have low mechanical and thermal properties compare to the broader set of hypothetical \gls*{rop} polymers that can potentially be made.

\subsection{Promising Polymers}

\begin{figure}
\centering
\includegraphics[width=\textwidth]{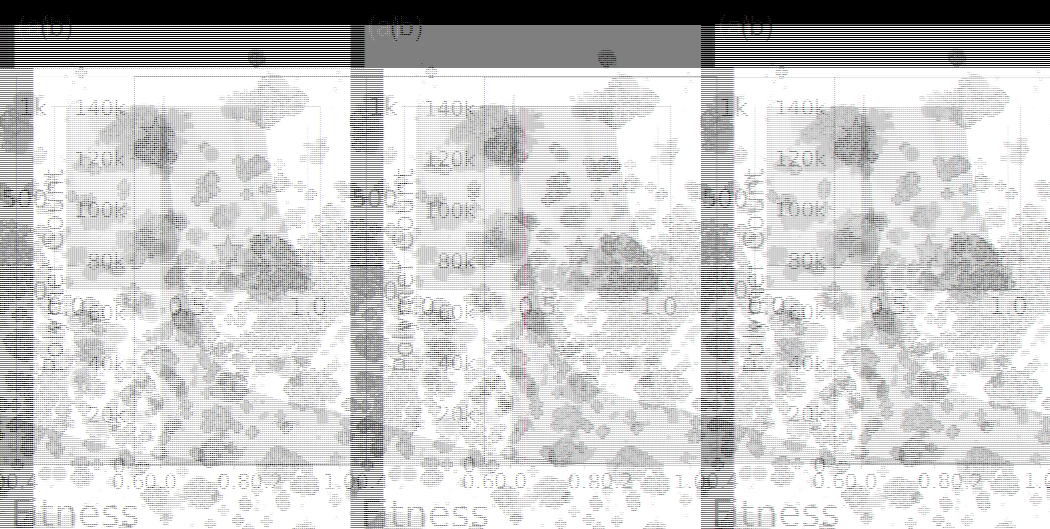}
\caption{(a) Histogram illustrating the distribution of fitness values among hypothetical polymers. (b) Uniform manifold approximation and projection (UMAP) visualization of high-dimensional polymer fingerprints condensed into two dimensions, with class colors matching (a) to indicate target-achieving polymers. The stars correspond to the bottom three polymers shown in \autoref{tab:tab2}.}
\label{fig:fig5}
\end{figure}

To determine which hypothetical polymers are most promising, we employed a three-step fitness function defined in \secref{methods:fitness}. This assessment multiplies scaled property predictions, each normalized between zero and one, to compute a polymer's fitness score. Polymers achieving a perfect fitness score of one met all predefined property requirements. The fitness distribution among the polymers and their classes is illustrated in the stacked bar plot in \autoref{fig:fig5} (a).

The multiplicative approach employed in our fitness assessment heavily penalizes polymers that significantly fall short of achieving a single property or moderately miss multiple targets. This is reflected in the right-skewed distribution of polymer classes, characterized by a sharp decline at 0, a gradual decrease between 0.2 and 0.4, and a more pronounced drop-off thereafter. Furthermore, due to the capping of values above the target at 1, a clustering of polymers at the maximum fitness value of 1 is observed.

To account for modeling errors, we established a fitness score threshold of 0.8 to identify promising polymer candidates. Despite generating millions of polymers, only 817 achieved all property targets, while 35,321 met the 0.8 cutoff, underscoring the intricate challenge of designing polymers with multiple desired properties. The successful polymers were predominantly thioethers, ethers, and amides, with smaller subsets of cycloalkenes, thionoesters, thioesters, esters, and a negligible number of thiocanes and carbonates. However, amides, esters, thionoesters and carbonates exhibited a higher frequency of meeting the target criteria, relative to their representation in the population, as shown in \autoref{tab:tab1}.

We visualized the high-dimensional polymer fingerprint, as described in \secref{methods:fingerprinting}, using a \gls*{umap}\cite{mcinnesUMAPUniformManifold2018}. Employing a cosine similarity metric, we configured the UMAP with 200 nearest neighbors and a minimum distance setting of 0.25, optimizing for a balance between local and global manifold structures and ensuring adequate spacing of data points \cite{bajuszWhyTanimotoIndex2015}. The result is shown in \autoref{fig:fig5} (b). Known ROP polymers are denoted by green circles, while known ROP polymers achieving the desired fitness score, discussed in \secref{validation}, are marked with colored stars. Hypothetical ROP polymers achieving the fitness score are represented by colored plus symbols, contrasting with the grey x's representing other hypothetical ROP polymers. The color scheme for stars and plus symbols in (b) corresponds to the class colors in (a), facilitating intuitive comparison.

The \gls*{umap} visualization unveils discernible clusters, particularly noticeable on the upper right side, where hypothetical polymers exhibit substantial deviations from current \gls*{rop} chemistries, hinting at uncharted realms with untapped potential. Furthermore, some promising polymers are situated near established \gls*{rop} domains, particularly in the left portion of the map. This proximity suggests the existence of novel candidate polymers that may be able to effectively leverage known \gls*{rop} reaction catalysts, solvents, and temperatures.

\begin{figure}
\centering
\includegraphics[width=\textwidth]{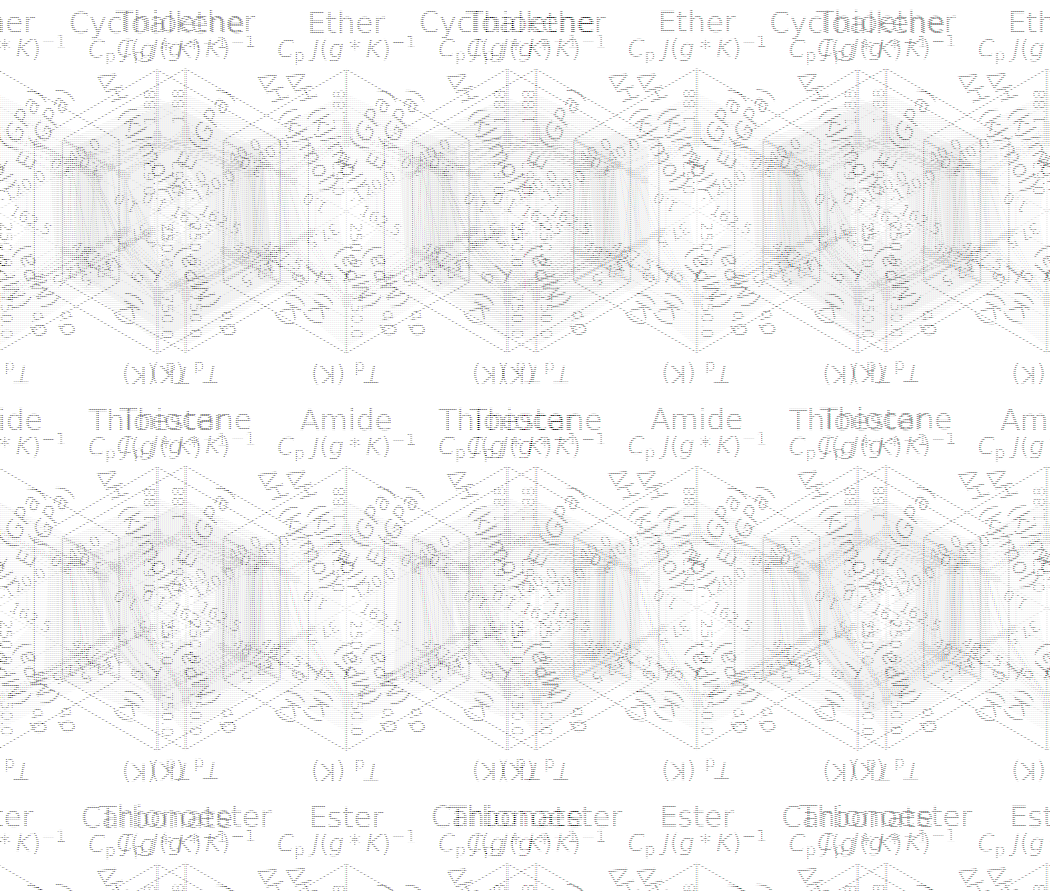}
\caption{Radar plots showcasing the property ranges of different polymer classes. Each shaded area delineates the minimum to maximum range of predicted properties within the respective polymer class, while the orange line denotes the targeted design region. Additionally, individual lines trace ten randomly chosen polymers that successfully meet the fitness objective ($>$0.8). The gold lines in the 'Amide`, 'Ester`, and 'Thiocane` charts delineate the predicted properties for the known ROP polymers within these classes that achieved the fitness objective.}
\label{fig:fig6}
\end{figure}

To comprehensively examine the distinct properties of the polymer classes, we devised radar plots for each category, as depicted in \autoref{fig:fig6}. Within each plot, the shaded region delineates the minimum to maximum range of predicted properties for the polymer class and the solid lines represent the properties for ten randomly chosen polymers achieving the fitness threshold within the class. Additionally, the predicted properties for known ROP polymer within a class are colored gold.

The plot reveals that the majority of polymers hover around or slightly exceed the target thresholds for each property. Notably, some polymers excel in specific properties like \gls*{E} or \gls*{tg}, while barely meeting the other targets. Conversely, a subset of polymers falls short of the target region for certain properties, particularly \gls*{dh}. However, these deviations are marginal, and therefore, the fitness function imposes only minimal penalties.

\autoref{tab:tab2} showcases a selected group of hypothetical polymers that achieved all targeted properties, along with their predicted properties. A common structural motif among all promising polymers is the presence of cyclic elements within their backbones or side chains, which is known to enhance polymer rigidity and consequently improve thermal stability, stiffness, and strength \cite{chungLiquidCrystallinePolymers2003, yuUnravelingSubstituentEffects2018}. In our dataset of known polymers, we see a statistically significant upward shift in the histograms of \gls*{tg}, \gls*{E}, and \gls*{tsb} for polymers with either aliphatic or aromatic rings. Moreover, there is a notable further shift towards higher values when these rings are incorporated within the backbone, as illustrated in supplementary Figure S1. For the hypothetical polymers, those with the ring in the backbone also tended to have higher \gls*{tg}.

\begin{table}
\caption{Predicted and Measured Properties of Promising ROP Polymer Candidates\tablefootnote{The first 12 candidates represent hypothetical polymers, while the last three have been synthesized and tested \cite{zhuSyntheticPolymerSystem2018a,cywarRedesignedHybridNylons2022}. The selection process for the hypothetical polymers was based on the length of the \gls*{smiles}, serving as a heuristic for polymer complexity, with the 12 shortest strings presented here. For synthesized polymers, the measured properties are shown in parentheses below the predicted values. The complete list list of hypothetical polymers, including their corresponding fitness values, is available in our data repository, referenced in \secref{Data}.}}
\label{tab:tab2}
\centering
\begin{tabular}{@{}c|c|cccccc@{}}
\toprule

\centering Class & Polymer & \makecell{\gls*{tg} \\ (K)} & \makecell{\gls*{td} \\ (K)} & \makecell{\gls*{E} \\ (GPa)} & \makecell{\gls*{tsb} \\ (MPa)} & \makecell{\gls*{dh} \\ (kJ/mol)} & \makecell{\gls*{cp} \\ (J(gK)$^{-1}$)} \\

\midrule

Amide & \includegraphics[width=1cm]{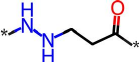} & 397     & 612      & 2.21     & 80.6     & -18.45   & 1.42\\

Ether & \includegraphics[width=1cm]{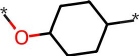} & 381      & 580      & 2.71     & 54.72    & -10.26   & 1.62\\

Ether & \includegraphics[width=1cm]{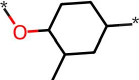} & 386      & 558      & 2.07     & 41.66    & -15.31   & 1.66\\

Ether & \includegraphics[width=1cm]{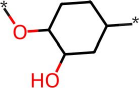} & 394      & 511      & 2.68     & 57.84    & -16.22   & 1.48\\

Ether & \includegraphics[width=1cm]{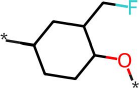} & 384      & 568      & 2.44     & 51.17    & -16.45   & 1.6\\

Amide & \includegraphics[width=1cm]{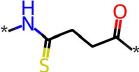} & 479      & 725      & 2.24     & 80.06    & -10.15   & 1.34\\

Ether & \includegraphics[width=1cm]{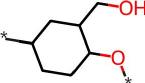} & 392      & 530      & 2.27     & 50.06    & -16.14   & 1.59\\

Thioether & \includegraphics[width=1cm]{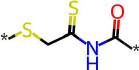} & 470      & 689      & 2.14     & 77.52    & -13.55   & 1.26\\

Ether & \includegraphics[width=1cm]{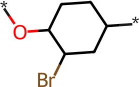} & 385      & 525      & 2.73     & 57.99    & -13.91   & 1.41\\

Ether & \includegraphics[width=1cm]{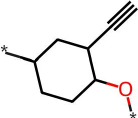} & 419      & 622      & 2.47     & 60.76    & -11.78   & 1.51\\

Ether & \includegraphics[width=1cm]{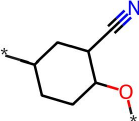} & 434      & 623      & 2.42     & 63.63    & -14.72   & 1.42\\

Ether & \includegraphics[width=1cm]{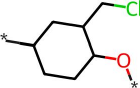} & 393      & 619      & 2.27     & 48.17    & -16.45   & 1.43\\

\makecell{Ester \\ (Known)} & \includegraphics[width=1cm]{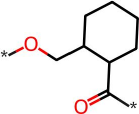} &  \makecell{393 \\ (322)}    & \makecell{595 \\ (613)} & \makecell{2.22 \\ (2.7)} & \makecell{47.24 \\ (54.7)} & \makecell{-19.63 \\ (-20)}    & 1.28     \\

\makecell{Amide \\ (Known)} & \includegraphics[width=1cm]{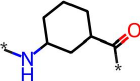} & \makecell{473 \\ (419-472)} & \makecell{646 \\ (630)}    & \makecell{2.02 \\ (2.28)} & 52.15 & \makecell{-12.48 \\ (-10)} & 1.37 \\

\makecell{Thiocane \\ (Known)} & \includegraphics[width=2cm]{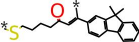} & \makecell{379 \\ (352)} & \makecell{613 \\ (563)}    & \makecell{1.93} & 48.1 & \makecell{1.92} & 1.37 \\

\bottomrule
\end{tabular}
\end{table}

Among the 35,321 selected polymers, only 416 lacked any ring structures entirely. Notably, 323 of these polymers featured alcohol, (thio)imide or amide moieties with no or short side chains. We hypothesize that the observed enhanced performance of these interesting cases may stem from the potential for hydrogen bonding within these structures. Indeed, hydrogen bonding has been recognized to augment mechanical properties by promoting inter-chain interactions that enhance polymer rigidity \cite{scheeljeNitrogenContainingPolymersDerived2023,vandersmanPredictionsGlassTransition2013,maRoleIntrinsicFactors2019,liEffectHydrogenBond2021,huangInsightsRoleHydrogen2021}, so long as intrachain bonding that can increase chain flexibility is avoided \cite{maRoleIntrinsicFactors2019}. Moreover, side chains are known to influence polymer properties significantly. Increasing the size of side chains typically leads to greater free volume and reduced packing density, factors known to lower both the \gls*{tg} and mechanical strength \cite{sugiyamaEffectsFlexibilityBranching2018,cowieEffectSideChain1990}.

Mirroring the stiffening trend observed in ring-containing polymers, we once again noted a rightward shift in the histograms depicting the property distributions of known polymers containing nitrogen, as shown in supplementary Figure S2. While the presence of amines and amides seemed to have little impact on mechanical properties, thermal properties saw a small effect. Furthermore, both categories of properties exhibited a statistically significant rightward shift when featuring an imide structure. In general, we observed that an increase in the number of hydrogen-bond acceptor atoms in the chain corresponded to an increase in both properties. Moreover, a distinct positive correlation between \gls*{tg} and \gls*{tsb} was discerned in our experimental dataset, as illustrated in Figure S3. This is likely because both properties are influenced by chain stiffness.

The downside to functional groups that induce hydrogen bonding, however, is their tendency to be inherently reactive. Chemical intuition dictates caution against the inclusion of amines, hydroxyls, carboxylic acids, and acidic methylene groups in the monomer structures. While these groups have the potential to enhance thermal and mechanical properties through hydrogen bonding, their high reactivity could lead to undesirable side reactions during synthesis, ultimately diminishing the likelihood of successful ring-opening polymerization. Filtering out molecules containing these functional groups yields a list of 6,477 promising polymer candidates.

\section{Synthetic Validation \& Design Guidelines}
\label{validation}

Our design process identified two previously known polymers as potential candidates to replace \gls*{ps} (these were not part of the training set used to create our non-enthalpy property prediction \gls*{ml} models). Significantly, both featured rings in their backbone structures, with one additionally incorporating an amide moiety. These are displayed in the last two rows of \autoref{tab:tab2}. The polymer lacking nitrogen was synthesized from a $\Gamma$-butyrolactone derivative and successfully met all target properties except for the \gls*{cp}, which was not determined, and the \gls*{tg}, which we predicted to be 393 K but was measured at 322 K. The deviation in \gls*{tg} could potentially be attributed to the polymer's low crystallization rate or variations in stereochemistry that are not adequately represented by the models \cite{zhuSyntheticPolymerSystem2018a}. The other values listed were a \gls*{dh} of -20 kJ/mol, a \gls*{td} of 613 K, a \gls*{E} of 2.7 GPa, and a \gls*{tsb} of 54.7 MPa, all slightly  higher than our model predictions, but close.

For the second polymer, experimental measurements revealed a \gls*{td} of 630 K and a \gls*{tg} spanning from 419 K to 472 K. Impressively, the polymer demonstrated outstanding chemical recycling capabilities, boasting a remarkable mass recovery rate of 93-98\%. Moreover, upon copolymerization with nylon 4, it achieved an impressive \gls*{E} value of 2.28 GPa. Although literature lacks information regarding its \gls*{cp} or \gls*{tsb}, the other properties were consistent with both our predictions and design targets \cite{cywarRedesignedHybridNylons2022}.

The discovery of these polymers marks a compelling initial validation of our informatics-based approach. Moreover, they exhibit chemical motifs akin to other polymers flagged by our models. This promising discovery bodes well for unveiling more mechanically and thermally resilient, chemically recyclable polymers (poised to serve as ideal substitutes for \gls*{ps}) within our catalog.

In addition to the above ``designed" candidates previously identified in the literature, we endeavored to synthesize the other promising, entirely new candidates. Focusing on the thiocane class, we selected one functionalized with an alkyne-containing, low-cost dimethyl fluorene, depicted at the end of \autoref{tab:tab2}. In selecting this candidate, we utilized polymerization efficacy heuristics; dimethyl fluorene lacked additional functional groups that could potentially disrupt the polymerization process. The polymer was synthesized according to the specifications outlined in Supplementary Section ``Thiocane Synthesis". Excitingly, the \gls*{td} of the synthesized polymer met our target, and while the \gls*{tg} did not fall precisely within our desired range, it reached 352 K—only 27 K lower than the predicted value of 379 K. This discrepancy between predicted and measured values may be attributed to the polymer's relatively low molecular weight (6 kDa), suggesting potential improvement with higher molecular weights based on the Flory-Fox equation.

Unfortunately, we encountered difficulties during the polymer synthesis. Low yield in monomer preparation posed an initial challenge, limiting our ability to produce sufficient quantities of monomer for polymerization. Additionally, we were unable to exceed a molecular weight of 6 kDa as a result of solubility problems, which hindered measurements of mechanical properties. Despite numerous adjustments to the polymerization solvent, these challenges persisted. Nonetheless, this experience serves as robust validation of the \gls*{vfs} technique and underscores the inherent complexity of translating polymer ``design" to polymer synthesis. Among thousands of designs, only a select few are predicted to achieve the requisite property targets, and fewer still prove viable for synthesis due to potential interference from certain functional groups in the reaction pathway. Even when a viable candidate shows initial promise, achieving high monomer yields and large polymer molecular weights remains a formidable task. Thus, although employing informatics approaches like may not lead to an immediate synthesizable hit, it can be instrumental in prioritize efforts.

A comprehensive catalog of the promising polymers, along with their predicted properties, precursor materials, and suggested synthesis routes, is provided in the supplementary materials. Typically, the most promising candidates exhibit the following characteristics:

\begin{enumerate}
\item Preference for the heterocycle being opened to have a size of 4, 5, 6, or 7. These comprised 6\%, 46\%, 35\%, and 11\% of the down-selected polymers, respectively.
\item Presence of rings within the backbone structure, observed in 64\% of down-selected polymers.
\item Inclusion of rings somewhere in the polymer structure (either the backbone or side groups), with 99.3\% of down-selected polymers meeting this criterion.
\item Absence of side chains or presence of short chains or those containing bulky rings.
\item Ability to hydrogen bond (however, care is needed to ensure the bonding group does not interfere with the polymerization chemistry).
\end{enumerate}

% =========================================================================== %
% Conclusion
\section{Conclusion}
\label{sec:conclusion}

In this study, informatics advancements were employed to generate over seven million hypothetical \gls*{rop} polymer candidates from a dataset comprising 30,272,000 known and commercially available molecules, with the aim of identifying recyclable alternatives to conventional plastics that are both thermally and mechanically durable. Our methodology yielded over 35,000 promising recyclable candidates demonstrating predicted mechanical and thermal durability close to \gls*{ps}. With input from polymer chemists, we further narrowed down this number to 6,477. Furthermore, we identified two known chemically recyclable polymers, not initially included in our thermal and mechanical property training datasets, yet meeting the criteria for \gls*{ps} substitutes alongside our hypothetical polymers, serving as validation for our informatics approach. Additionally, we synthesized a novel thiocane polymer designed using our methodology with enhanced thermal properties. Investigation of all promising candidates revealed that suitable substitutes for \gls*{ps} containers shared similar chemical characteristics, notably involving the opening of heterocycles ranging from 4 to 7 atoms, presence of a ring within the polymer backbone, the absence of side chains, or the incorporation of bulky constituents within them. Additionally, numerous polymers contained atoms conducive to hydrogen bonding.

While \gls*{vfs} shows promise in generating synthetically feasible polymer candidates, only a minute fraction ($<$ 0.5\%) of the polymers produced meet all desired properties, suggesting avenues for accelerated discovery, potentially through integrating genetic algorithms \cite{kimPolymerDesignUsing2021a,kernDesignPolymersEnergy2021}. This will be crucial if expansion to copolymers, which show promise for recyclable designs, is pursued as the number of hypothetical polymers will quickly explode \cite{kuennethCopolymerInformaticsMultitask2021, shuklaPolymerInformaticsHomopolymers2024, zhangRecyclablePolyesterLibrary2023, illyAlternatingCopolymerizationBiobased2021, kumarAlternatingSulfoneCopolymers2015, takebayashiCationicRingopeningCopolymerization2023}. 

The reliance on ML models warrants caution, especially during extrapolation to unseen chemical spaces. Enhancements to these models, particularly by using multi-fidelity methods that blend lower fidelity simulation data with experimental observations, will be crucial for extrapolating to unexplored chemical regions \cite{tolandAcceleratedSchemePredict2023, venkatramPredictingCrystallizationTendency2020, patraMultifidelityInformationfusionApproach2020, batraMultifidelityInformationFusion2019}. Regardless, informatics approaches, when executed synergistically and iteratively with physical experiments, have the potential to rapidly expedite the discovery of novel and beneficial materials.

% =========================================================================== %
% Methods
\section{Methods}
\label{methods}

\subsection{Database}
\label{methods:db}

A PostgreSQL version 12.17 on an Ubuntu 20.04.6 LTS system was used to store all molecule, reaction, and polymer data. The full details of the database schema can be found in Supplementary Section ``Database Schema".

\subsubsection{Molecule Data}

The molecule section of the database features a ``category" column that delineates between known and hypothetical molecules. Known molecules, crucial for generating synthetically accessible polymers, are drawn from five distinct subsets: ZINC15, ChemBL, compounds sourced from literature, an eMolecules database dump from December 19th, 2020, and data from a VWR database, harvested through a tailored webscraper (details available in Supplementary Section ``Webscraper")\cite{sterlingZINC15Ligand2015,gaultonChEMBLLargescaleBioactivity2012,VwrEMoleculesCom,emoleculesBuyResearchCompounds}. Molecules from the ZINC15 and ChemBL datasets are reportedly ``commercially available," although in reality, they can be challenging to procure. Those sourced from literature theoretically can be synthesized based on the referenced procedures. Molecules in the eMolecules database are likely available commercially, though without assurance. However, each eMolecules ID is catalogued, facilitating easy cross referencing from the eMolecules site. Entries from the VWR database, covering the years 2023-2024, are also presumed to be commercially accessible.

\subsection{Predictive Models for Polymer Properties}
\label{methods:models}
The prediction of polymer properties outlined in \autoref{fig:fig1} relied on two distinct types of machine learning (ML) models. The first type employs \gls*{gpr} to forecast the \gls*{dh} for \gls*{rop} polymers. The second type uses \gls*{mtnn}s to predict the remaining polymer properties.

\subsubsection{Enthalpy of Ring-Opening Polymerization: Gaussian Process Regression (GPR)}
The \gls*{gpr} model was trained using a dataset consisting of experimental \gls*{dh} values from \gls*{rop} polymers, supplemented with \gls*{dft} data. This hybrid dataset effectively predicts experimental \gls*{dh} values for \gls*{rop} polymers \cite{tolandAcceleratedSchemePredict2023, tranRecyclablePolymersRingOpening2022}. The choice of \gls*{gpr} was motivated by its proven efficacy in accurately predicting polymer properties, particularly with small datasets.

\subsubsection{Multitask Neural Network (MTNN)}
Multiple \gls*{mtnn} models were trained using a diverse array of homopolymer and copolymer property data. These models were trained based on distinct classes of correlated properties, including thermal, mechanical, gas permeability, thermodynamic \& physical, electronic, and optical \& dielectric properties. The focus of this study, depicted in \autoref{fig:fig1} (a), primarily involved thermal, mechanical, and thermodynamic properties. For comprehensive details on the training and testing methodologies, refer to the original \gls*{mtnn} paper \cite{kuennethPolyBERTChemicalLanguage2023}.

\subsection{Fingerprinting Techniques}
\label{methods:fingerprinting}
Both the \gls*{gpr} model and \gls*{mtnn} models rely on the molecular features of polymers for predictions, with the \gls*{gpr} model also requiring monomer molecular features.

\subsubsection{Polymer Fingerprinting}
Polymer fingerprinting entails three hierarchical levels of descriptors. The initial level quantifies atomic triplets (e.g., H1–C4–H1, denoting two one-fold coordinated hydrogens and a four-fold coordinated carbon). The subsequent level encapsulates predefined chemical building blocks (e.g., –C6H4–, –CH2–, –C(= O)–). The third level encompasses Quantitative Structure-Property Relationship (QSPR) descriptors, incorporating molecular features like molecular quantum numbers and molecular connectivity chi indices, alongside other descriptors such as non-hydrogen atom count and molecular weight. These features are then normalized by the number of atoms in the polymer. Additional details can be found at \cite{kimPolymerGenomeDataPowered2018, doantranMachinelearningPredictionsPolymer2020a}.

For copolymers, each feature is derived through a linear combination of homopolymer features, with weights determined by the fraction of the respective homopolymer in the copolymer \cite{kuennethCopolymerInformaticsMultitask2021, kuennethPolyBERTChemicalLanguage2023}.

\subsubsection{Molecule Fingerprinting}
Similar to polymer fingerprinting, molecule fingerprinting involves three hierarchical levels of descriptors. However, unlike polymers, these descriptors are not normalized based on the number of atoms. Additionally, certain descriptors like the length of the longest side chain are disregarded. Crucially, \gls*{rop}-specific features such as the size of the ring being opened and the valence electron differences in the broken bond of the ring are integrated into the fingerprinting process \cite{tolandAcceleratedSchemePredict2023, kernSolventSelectionPolymers2022}.

\subsection{Fitness Assessment}
\label{methods:fitness}

The evaluation of hypothetical polymers against the screening criteria depicted in \autoref{fig:fig1} (b) and (c) entailed a three-step procedure:

\begin{enumerate}
    \item \textbf{Enthalpy Transformation}: The target range for \gls*{dh} was set between -10 kJ/mol and -20 kJ/mol. To ensure uniform weighting between values less than -20 kJ/mol and greater than -10 kJ/mol, \gls*{dh} values were adjusted using \autoref{eq:1}:
    \begin{equation}
    \label{eq:1}
        \Delta H_{i}^{t}= 
        \begin{cases}
            \Delta H_{i} + 30,& \text{if } \Delta H_{i} < -20\\
            10, & \text{if } -20 \geq \Delta H_{i} \leq -10\\
            \Delta H_{i} * -1, & \Delta H_{i} > -10
        \end{cases}
    \end{equation}
    Here, $\Delta H_{i}^{t}$ stands for the transformed \gls*{dh} value for polymer $i$ and $\Delta H_{i}$ stands for the predicted \gls*{dh} value for polymer $i$.
    
    \item \textbf{Clipping of Predicted Properties}: Predicted properties exceeding the target thresholds were clipped to the targets. This process is mathematically represented by \autoref{eq:2}:
    \begin{equation}
    \label{eq:2}
        k'_{i} = min(k_{i}, k_{target})
    \end{equation}

    In the equation, $k_{i}$ represents the predicted value of property $k$ for polymer $i$, where $k$ can be any of the following: \gls*{tg}, \gls*{td}, \gls*{tsb}, \gls*{E}, \gls*{cp}, or \gls*{dh}. $k_{target}$ denotes the minimum threshold that a property must meet, as defined in \autoref{fig:fig1} (b) and (c). $k'_{i}$ signifies the clipped predicted value of polymer $i$: it remains unchanged if the value is below the target, or becomes the target if it exceeds it. This calculation prioritizes polymers that satisfy all criteria over those that excel in only a few aspects.

    \item \textbf{Normalization and Fitness Calculation}: The adjusted property values were normalized within the range of 0 to 1 using a MinMaxScaler. A composite fitness value for each polymer was then calculated by multiplying these normalized properties, as described in \autoref{eq:3}:
    \begin{equation}
    \label{eq:3}
        \theta_{i} = \prod_{k=T_{g}, T_{d}, E, \sigma_{b}, C_{p}, \Delta H^{t}}  \frac{k'_{i} - k'_{min}}{k_{target} - k'_{min}}
    \end{equation}

    Here, $k'_{min}$ represents the minimum clipped predicted value for all polymers in the dataset. $\theta_{i}$ represents the fitness score for polymer $i$.
\end{enumerate}

\section*{Acknowledgements}
This work is supported by the Office of Naval Research through a Multidisciplinary University Research Initiative (MURI) Grant (N00014-20-1-2586). J.K. gratefully acknowledges support through the National Defense Science and Engineering (NDSEG) Fellowship Program from the Department of Defense (DoD).

\section*{Data Availability}
\label{Data}
All relevant resources, including polymer designs, detailed reaction procedures, and accompanying code for generating polymers from monomers and reaction procedures, are publicly available in our \href{https://github.com/Ramprasad-Group/polyVERSE/tree/main/Virtual-Polymer/VFS/ROP/packaging_replacements}{polyVERSE repository on GitHub}.

\clearpage

\bibliography{references}

\clearpage
\printglossary[type=\acronymtype]

\end{document}